\documentclass[preprintnumbers,amsmath,amssymb,showpacs]{revtex4}
\usepackage{subfigure}
\usepackage{dsfont}
\usepackage[landscape,papersize={297.1mm,210mm},left=1.9cm,right=1.6cm,top=2.7cm,bottom=2.8cm]{geometry}
\usepackage[                   
            pdfstartview=FitH,
            colorlinks, 
            pdfborder=001,   
            linkcolor=blue,
            anchorcolor=blue,
            citecolor=blue,
            urlcolor=blue
            ]{hyperref}
\usepackage{graphicx}
\usepackage{epstopdf}
\usepackage{dcolumn}
\usepackage{bm}

\begin{document}
\title{Detection of the quantum states containing at most $k-1$ unentangled particles}

\author{Yan Hong}
 \affiliation {School of Mathematics and Science, Hebei GEO University, Shijiazhuang 050031,  China}

\author{Xianfei Qi}
 \affiliation {School of Mathematics and Statistics, Shangqiu Normal University, Shangqiu 476000,  China}

\author{Ting Gao}
\email{gaoting@hebtu.edu.cn} \affiliation {School of Mathematical Sciences, Hebei Normal University, Shijiazhuang 050024,  China}

\author{Fengli Yan}
\email{flyan@hebtu.edu.cn} \affiliation {College of Physics, Hebei Normal University, Shijiazhuang 050024,  China}

\begin{abstract}
There are many different classifications of entanglement for multipartite quantum systems, one of which
is based on the number of unentangled particles. In this paper, we mainly study the quantum states containing at most $k-1$ unentangled particles and provide several entanglement criteria based on  different forms of inequalities which can  both identify quantum states containing at most $k-1$ unentangled particles. We show that these criteria are more effective for some states by concrete examples.\\


\end{abstract}


\maketitle

\section{Introduction}
As a fundamental concept of quantum theory, quantum entanglement plays a crucial role in quantum information processing \cite{RMP81.865}. It has been successfully identified as a key ingredient for a wide range of applications, such as quantum cryptography \cite{PRL67.661},  quantum dense coding \cite{PRL69.2881}, quantum teleportation \cite{PRL70.1895,EPL84.50001}, factoring \cite{SIAM1997}, and quantum computation \cite{Nature2000}.

One of the significant problems in the study of quantum entanglement theory is to decide whether a quantum state is entangled or not. For bipartite systems, quantum states consists of separable states and entangled states. Many well-known separability criteria have been proposed to distinguish separable from entangled states \cite{PR474.1,NRP2019}. In multipartite case, the classification of quantum states is much more complicated due to the complex structure of multipartite quantum states. A reasonable way of classification is based on the number of partitions that are separable. According to that, $N$-partite quantum states can be divided into $k$-separable states and $k$-nonseparable states, with $2\leqslant k\leqslant N$. The detection of $k$-nonseparability has been investigated extensively, many efficient criteria \cite{QIC2008,QIC2010,PRA82.062113,EPL104.20007,PRA91.042313,SR5.13138,PRA93.042310,SciChina2017} and computable measures \cite{PRA68.042307,PRL93.230501,PRA83.062325,PRA86.062323,PRL112.180501} have been presented. Different from the above classification, $N$-partite quantum states can also be divided into $k$-producible states and $(k+1)$-partite entangled states by consideration of the number of partitions that are entangled. It is worth noting that the $(k+1)$-partite entanglement and the $k$-nonseparability are two different concepts involving the partitions of subsystem in $N$-partite quantum systems, and they are equivalent only in some special cases.

In this paper we focus on another characterization of multipartite quantum states which is based on the number of unentangled particles. We first present the definition of quantum states containing at least $k$ unentangled particles, and then derive several criteria to identify quantum states containing at most $k-1$ unentangled particles by using some well-known inequalities. Several specific examples illustrate  the advantage of our results in detecting quantum states containing at most $k-1$ unentangled particles.

The organization of this article is as follows: In Sec.~\uppercase\expandafter{\romannumeral 2} we review basic knowledge which will be used in the rest of the paper. In Sec.~\uppercase\expandafter{\romannumeral 3}, we provide our central results,  several criteria that can effectively detect quantum states containing at most $k-1$ unentangled particles, and then their strengths are exhibited by several examples. Finally, a brief summary is given in Sec.~\uppercase\expandafter{\romannumeral 4}.

\section{Preliminaries}
In this section, we introduce preliminary knowledge used in this paper. We consider a multiparticle quantum system with state space $\mathcal{H}=\mathcal{H}_{1} \otimes \mathcal{H}_{2} \otimes \cdots \otimes \mathcal{H}_{N}$, where $\mathcal{H}_{i}~(i=1,2,\ldots,N)$ denote $d_{i}$-dimensional Hilbert spaces. For convenience, we introduce the following concepts. An $N$-partite pure state $|\psi\rangle\in \mathcal{H}_1\otimes \mathcal{H}_2\otimes\cdots\otimes \mathcal{H}_N$ contains  $k$ unentangled particles, if there is $k+1$ partition $\gamma_1|\gamma_2|\cdots|\gamma_{k+1}$ such that
$$|\psi\rangle=\bigotimes\limits_{l=1}^{k+1}|\phi_l\rangle_{\gamma_l},$$
where $|\phi_l\rangle_{\gamma_l}$ is single-partite state for $1\leqslant l\leqslant k$, while $|\phi_{k+1}\rangle_{\gamma_{k+1}}$ is a ($N-k$)-particle state.
A mixed state $\rho$ contains at least $k$ unentangled particles, if it can be written as
$$\rho=\sum_{j}p_{j}|\psi^{(j)}\rangle\langle\psi^{(j)}|,$$
where $p_{j}>0$ with $\sum_{}p_{j}=1$, and $|\psi^{(j)}\rangle$ is the pure state containing   $m_j$ unentangled particles with $m_{j}\geqslant k$ \cite{Goth2005,Goth2009,Goth2012}. Otherwise we say $\rho$ contains at most $k-1$ unentangled particles.

For $N$-partite quantum system $\mathcal{H}_1\otimes \mathcal{H}_2\otimes\cdots\otimes \mathcal{H}_N$, let
\begin{equation}\label{tr}
\begin{array}{rl}
\langle\bigotimes\limits_{i=1}^{N}A_{i}B_{i}\rangle_\rho=\textrm{tr}\big((\bigotimes\limits_{i=1}^NA_iB_i)\rho\big)
 \end{array}
\end{equation}
 where $\rho$ is the quantum state in $\mathcal{H}_1\otimes \mathcal{H}_2\otimes\cdots\otimes \mathcal{H}_N$, $A_i, B_i$ are  operators acting on the $i$-th subsystem $\mathcal{H}_i$, and ``tr" stands for trace operation.

Inequalities play an important role in quantum information theory. In the following, we list some inequalities that will be used throughout the paper.

Absolute value inequality:
\begin{equation}\label{AVI}
\begin{array}{rl}
\left|\sum\limits_{i=1}^{n}a_{i}\right|\leq\sum\limits_{i=1}^{n}\left|a_i\right|.
\end{array}
\end{equation}

Cauchy-Schwarz inequality:
\begin{equation}\label{CSI1}
\begin{array}{rl}
|\langle x|y\rangle|^2\leq\langle x|x\rangle\langle y|y\rangle,
\end{array}
\end{equation}

\begin{equation}\label{CSI2}
\begin{array}{rl}
\left(\sum\limits_{i=1}^{n}a_{i}b_{i}\right)^2\leq\left(\sum\limits_{i=1}^{n}a_i^2\right)\left(\sum\limits_{i=1}^{n}b_i^2\right).
\end{array}
\end{equation}

Extending the Cauchy-Schwarz inequality is an important inequality known as the  H$\ddot{\textrm{o}}$lder inequality:
 \begin{equation}\label{HI}
\begin{array}{rl}
\sum\limits_{i=1}^n|a_ib_i|\leq\left(\sum\limits_{i=1}^n|a_i|^p\right)^{\frac{1}{p}}\left(\sum\limits_{i=1}^n|b_i|^q\right)^{\frac{1}{q}},
\end{array}
\end{equation}
 where $p,q>1,\dfrac{1}{p}+\dfrac{1}{q}=1$.

\section{Main Results}
Now let us state our criteria identifying  quantum states containing at most $k-1$ unentangled particles for arbitrary dimensional multipartite quantum systems.

\textbf{Theorem 1.}~If an $N$-partite quantum states $\rho$ contains at least $k$ unentangled particles for $1\leqslant k\leqslant N-1$,  then it  satisfies
\begin{equation}\label{k-pro-p}
\begin{array}{rl}
&\left|\langle\bigotimes\limits_{i=1}^NA_iB_i\rangle_\rho\right|\\
\leq&\sum\limits_{\gamma}\Big(\prod\limits_{l=1}^{k+1}\big\langle(\bigotimes\limits_{i\in\gamma_l}A_iA_i^\dagger)\otimes(\bigotimes\limits_{i\notin\gamma_l}B_i^\dagger B_i)\big\rangle_{\rho}
\big\langle(\bigotimes\limits_{i\in\gamma_l}B_i^\dagger B_i)\otimes(\bigotimes\limits_{i\notin\gamma_l}A_i A_i^\dagger)\big\rangle_{\rho}\Big)^{\frac{1}{2k+2}}
\end{array}
\end{equation}
where $A_i, B_i$ are operators acting on the $i$-th subsystem, the sum runs over all possible partitions $\{\gamma|\gamma=\gamma_1|\gamma_2|\cdots|\gamma_{k+1}\}$ of $N$ particles in which the number of particles in $\gamma_l$ is  1 for $1\leq l\leq k$ and is $N-k$ for  $i=k+1$ .
If $\rho$ violates inequality (\ref{k-pro-p}), then it contains at most $k-1$ unentangled particles.

\emph{Proof.} First, we consider  pure state $\rho=|\psi\rangle\langle\psi|$ containing $k$ unentangled particles. Suppose that the pure state
$|\psi\rangle=\bigotimes\limits_{l=1}^{k+1}|\psi_l\rangle_{\gamma_l}$ under the partition $\gamma_1|\gamma_2|\cdots|\gamma_{k+1}$,
where $\gamma_l$ contains one particle for $1\leq l\leq k$, and $\gamma_{k+1}$ contains $N-k$ particles.
 Then for any subsystems $\gamma_l$, we have
$$\begin{array}{ll}
&\left|\langle\bigotimes\limits_{i=1}^NA_iB_i\rangle_\rho\right|\\
=&\sqrt{\langle\bigotimes\limits_{i=1}^NA_iB_i\rangle_\rho\langle\bigotimes\limits_{i=1}^NB_i^\dagger A_i^\dagger\rangle_\rho}\\
=&\sqrt{\prod\limits_{t=1}^{k+1}\langle\bigotimes\limits_{i\in\gamma_t}A_iB_i\rangle_{\rho_{\gamma_t}}
\prod\limits_{t=1}^{k+1}\langle\bigotimes\limits_{i\in\gamma_t}B_i^\dagger A_i^\dagger\rangle_{\rho_{\gamma_t}}}\\
\leq&\sqrt{\prod\limits_{t=1}^{k+1}\langle\bigotimes\limits_{i\in\gamma_t}A_i A_i^\dagger\rangle_{\rho_{\gamma_t}}
\prod\limits_{t=1}^{k+1}\langle\bigotimes\limits_{i\in\gamma_t}B_i^\dagger B_i\rangle_{\rho_{\gamma_t}}}\\
=&\sqrt{\langle\bigotimes\limits_{i\in\gamma_l}A_iA_i^\dagger\rangle_{\rho_{\gamma_l}}
\prod\limits_{t\neq l}\langle\bigotimes\limits_{i\in\gamma_t}A_i A_i^\dagger\rangle_{\rho_{\gamma_t}}
\langle\bigotimes\limits_{i\in\gamma_l}B_i^\dagger B_i\rangle_{\rho_{\gamma_l}}
\prod\limits_{t\neq l}\langle\bigotimes\limits_{i\in\gamma_t}B_i^\dagger B_i\rangle_{\rho_{\gamma_t}}}\\
=&\sqrt{\langle(\bigotimes\limits_{i\in\gamma_l}A_iA_i^\dagger)\otimes(\bigotimes\limits_{i\notin\gamma_l}B_i^\dagger B_i)\rangle_\rho
\langle(\bigotimes\limits_{i\in\gamma_l}B_i^\dagger B_i)\otimes(\bigotimes\limits_{i\notin\gamma_l}A_i A_i^\dagger)\rangle_\rho},
\end{array}$$
where $\rho_{\gamma_t}=|\psi_t\rangle_{\gamma_t}\langle\psi_t|$. Here we have used the Cauchy-Schwarz inequality (\ref{CSI1}). Thus,

\begin{equation*}\begin{array}{ll}
&\left|\langle\bigotimes\limits_{i=1}^NA_iB_i\rangle_\rho\right|\\
\leq&\Big(\prod\limits_{l=1}^{k+1}\sqrt{\big\langle(\bigotimes\limits_{i\in\gamma_l}A_iA_i^\dagger)\otimes(\bigotimes\limits_{i\notin\gamma_l}B_i^\dagger B_i)\big\rangle_\rho
\big\langle(\bigotimes\limits_{i\in\gamma_l}B_i^\dagger B_i)\otimes(\bigotimes\limits_{i\notin\gamma_l}A_i A_i^\dagger)\big\rangle_\rho}\Big)^{\frac{1}{{k+1}}}\\
\leq&\sum\limits_{\gamma}\Big(\prod\limits_{l=1}^{k+1}\sqrt{\big\langle(\bigotimes\limits_{i\in\gamma_l}A_iA_i^\dagger)\otimes(\bigotimes\limits_{i\notin\gamma_l}B_i^\dagger B_i)\big\rangle_\rho
\big\langle(\bigotimes\limits_{i\in\gamma_l}B_i^\dagger B_i)\otimes(\bigotimes\limits_{i\notin\gamma_l}A_i A_i^\dagger)\big\rangle_\rho}\Big)^{\frac{1}{{k+1}}}\\
=&\sum\limits_{\gamma}\Big(\prod\limits_{l=1}^{k+1}\big\langle(\bigotimes\limits_{i\in\gamma_l}A_iA_i^\dagger)\otimes(\bigotimes\limits_{i\notin\gamma_l}B_i^\dagger B_i)\big\rangle_{\rho}
\big\langle(\bigotimes\limits_{i\in\gamma_l}B_i^\dagger B_i)\otimes(\bigotimes\limits_{i\notin\gamma_l}A_i A_i^\dagger\big\rangle_{\rho}\Big)^{\frac{1}{2k+2}}.
\end{array}\end{equation*}
It shows that inequality (\ref{k-pro-p}) is right for pure state containing $k$ unentangled particles.

Now, we consider the case of mixed state. Suppose $\rho=\sum\limits_{j}p_{j}\rho_{j}$ is a mixed state with pure states $\rho_j$ containing at least $k$ unentangled particles, then
$$\begin{array}{ll}
&\left|\langle\bigotimes\limits_{i=1}^NA_iB_i\rangle_\rho\right|\\
\leq&\sum\limits_jp_j\left|\langle\bigotimes\limits_{i=1}^NA_iB_i\rangle_{\rho_j}\right|\\
\leq&\sum\limits_jp_j\sum\limits_{\gamma}\Big(\prod\limits_{l=1}^{k+1}\big\langle(\bigotimes\limits_{i\in\gamma_l}A_iA_i^\dagger)\otimes(\bigotimes\limits_{i\notin\gamma_l}B_i^\dagger B_i)\big\rangle_{\rho_j}
\big\langle(\bigotimes\limits_{i\in\gamma_l}B_i^\dagger B_i)\otimes(\bigotimes\limits_{i\notin\gamma_l}A_i A_i^\dagger)\big\rangle_{\rho_j}\Big)^{\frac{1}{{2k+2}}}\\
\leq&\sum\limits_{\gamma}\Big(\prod\limits_{l=1}^{k+1}\sum\limits_jp_j\big\langle(\bigotimes\limits_{i\in\gamma_l}A_iA_i^\dagger)\otimes(\bigotimes\limits_{i\notin\gamma_l}B_i^\dagger B_i)\big\rangle_{\rho_j}^{\frac{1}{2}}
\big\langle(\bigotimes\limits_{i\in\gamma_l}B_i^\dagger B_i)\otimes(\bigotimes\limits_{i\notin\gamma_l}A_i A_i^\dagger)\big\rangle_{\rho_j}^{\frac{1}{2}}\Big)^{\frac{1}{{k+1}}}\\
\leq&\sum\limits_{\gamma}\Big(\prod\limits_{l=1}^{k+1}\big\langle(\bigotimes\limits_{i\in\gamma_l}A_iA_i^\dagger)\otimes(\bigotimes\limits_{i\notin\gamma_l}B_i^\dagger B_i)\big\rangle_{\rho}
\big\langle(\bigotimes\limits_{i\in\gamma_l}B_i^\dagger B_i)\otimes(\bigotimes\limits_{i\notin\gamma_l}A_i A_i^\dagger)\big\rangle_{\rho}\Big)^{\frac{1}{2{k+2}}}
\end{array},$$
where we have used the absolute value inequality (\ref{AVI}), inequality (\ref{k-pro-p}) for pure states, the H$\ddot{\textrm{o}}$lder inequality (\ref{HI}) and Cauchy-Schwarz inequality (\ref{CSI2}). The proof is complete.

\textbf{Theorem 2.} If an $N$-partite quantum states $\rho$ is fully separable state (that is, it is  the quantum state containing at least $N-1$ unentangled particles), then we have
\begin{equation}\label{k}
\begin{array}{rl}
\left|\langle\bigotimes\limits_{i=1}^NA_iB_i\rangle_\rho\right|
\leq&\sqrt{\big\langle\bigotimes\limits_{i=1}^NA_iA_i^\dagger\big\rangle_{\rho}
\big\langle\bigotimes\limits_{i=1}^NB_i^\dagger B_i\big\rangle_{\rho}}.
\end{array}
\end{equation}
Of course, $\rho$ is entangled if it violates the inequality (\ref{k}).

\emph{Proof.} Suppose that
$|\psi\rangle=\bigotimes\limits_{i=1}^{N}|\psi_{i}\rangle$ is fully separable pure state, we have
$$\begin{array}{ll}
&\left|\langle\bigotimes\limits_{i=1}^NA_iB_i\rangle_{|\psi\rangle}\right|^2\\
=&\langle\bigotimes\limits_{i=1}^NA_iB_i\rangle_{|\psi\rangle}\langle\bigotimes\limits_{i=1}^NB_i^\dagger A_i^\dagger\rangle_{|\psi\rangle}\\
=&\prod\limits_{i=1}^{N}\langle A_iB_i\rangle_{|\psi_{i}\rangle}
\prod\limits_{l=1}^{N}\langle B_i^\dagger A_i^\dagger\rangle_{{|\psi_{i}\rangle}}\\
\leq&\prod\limits_{i=1}^{N}\sqrt{\langle A_iA^\dagger_i\rangle_{|\psi_{i}\rangle}\langle B_i^\dagger B_i\rangle_{{|\psi_{i}\rangle}}}
\prod\limits_{i=1}^{N}\sqrt{\langle B_i^\dagger B_i\rangle_{{|\psi_{i}\rangle}}\langle A_iA^\dagger_i\rangle_{|\psi_{i}\rangle}}\\
=&\prod\limits_{i=1}^{N}\langle A_iA^\dagger_i\rangle_{|\psi_{i}\rangle}\langle B_i^\dagger B_i\rangle_{{|\psi_{i}\rangle}}\\
=&\big\langle\bigotimes\limits_{i=1}^NA_iA_i^\dagger\big\rangle_{\rho}
\big\langle\bigotimes\limits_{i=1}^NB_i^\dagger B_i\big\rangle_{\rho},
\end{array}$$
where we have used the Cauchy-Schwarz inequality (\ref{CSI1}). Hence, inequality (\ref{k}) holds for fully separable pure state. Similar to Theorem 1, we can find that (\ref{k}) is guaranteed for fully separable mixed states.

Here we need to point out that Theorem 2 is inequality (44) of Ref.\cite{Wolk2014}.

\textbf{Theorem 3.}~For any $N$-partite density matrix acting on Hilbert space $\rho\in \mathcal{H}_1\otimes \mathcal{H}_2\otimes\cdots\otimes \mathcal{H}_N$
 containing at least $k$ unentangled particles, where $1\leq k\leq N-2$, we have
\begin{equation}\label{k-pro-s}
\begin{array}{rl}
\sum\limits_{m\neq n}\left|\langle U_m(\bigotimes\limits_{i=1}^NA_iA_i^\dagger) U_n^\dagger\rangle_\rho\right|
\leq\sum\limits_{m\neq n}\sqrt{\langle \bigotimes\limits_{i=1}^NA_iA_i^\dagger\rangle_\rho\langle U_mU_n(\bigotimes\limits_{i=1}^NA_iA_i^\dagger)U_n^\dagger U_m^\dagger\rangle_\rho}+(N-k-1)\sum\limits_m\langle U_m(\bigotimes\limits_{i=1}^NA_iA_i^\dagger) U_m^\dagger\rangle_\rho
\end{array}
\end{equation}
where  $A_i$ are any operator of the subsystem $\mathcal{H}_i$, and $U_m=\textbf{1}_1\otimes\cdots\otimes\textbf{1}_{m-1}\otimes u_m\otimes\textbf{1}_{m+1}\otimes\cdots\otimes\textbf{1}_N$  with $u_m$ being any operator of the subsystem $\mathcal{H}_m$ and $\textbf{1}_j$ being identity matrix of the subsystem $\mathcal{H}_j$. If $\rho$ violates (\ref{k-pro-s}), then it contains at most $k-1$ unentangled particles.

\emph{Proof.} We begin with pure state. Suppose that the pure state $|\psi\rangle$ contains  $k$ unentangled particles, then there is a  partition  $\gamma_1|\cdots|\gamma_{k+1}$ with  $\gamma_l$ containing one particle for $1\leq l\leq k$, and $\gamma_{k+1}$ containing  $N-k$ particles,
it can be written as $|\psi\rangle=\bigotimes\limits_{l=1}^{k+1}|\psi_l\rangle_{\gamma_l}$. When $m,n$ in $\gamma_{k+1}$,  we can obtain
\begin{equation}\label{sandt}
\begin{array}{rl}
&\left|\langle U_m(\bigotimes\limits_{i=1}^NA_iA_i^\dagger) U_n^\dagger\rangle_\rho\right|\\
\leq&\sqrt{\langle U_m(\bigotimes\limits_{i=1}^NA_iA_i^\dagger) U_m^\dagger\rangle_\rho\langle U_n(\bigotimes\limits_{i=1}^NA_iA_i^\dagger) U_n^\dagger\rangle_\rho}\\
\leq&\dfrac{\langle U_m(\bigotimes\limits_{i=1}^NA_iA_i^\dagger) U_m^\dagger\rangle_\rho+\langle U_n(\bigotimes\limits_{i=1}^NA_iA_i^\dagger) U_n^\dagger\rangle_\rho}{2},
\end{array}
\end{equation}
where first inequality holds because of Cauchy-Schwarz inequality (\ref{CSI1}) and second inequality follows from mean inequality.
When $m\in \gamma_l$,  $n\in \gamma_{l'}$ and $l\neq l'$, we have
\begin{equation}\label{sort}
\begin{array}{rl}
\left|\langle U_m(\bigotimes\limits_{i=1}^NA_iA_i^\dagger) U_n^\dagger\rangle_\rho\right|
\leq\sqrt{\langle(\bigotimes\limits_{i=1}^NA_iA_i^\dagger)\rangle_\rho\langle U_mU_n(\bigotimes\limits_{i=1}^NA_iA_i^\dagger)U_n^\dagger U_m^\dagger\rangle_\rho}
\end{array}
\end{equation}
by Cauchy-Schwarz inequality (\ref{CSI1}).

Combining (\ref{sandt}) and (\ref{sort}) leads to
$$\begin{array}{ll}
&\sum\limits_{m\neq n}\left|\langle U_m(\bigotimes\limits_{i=1}^NA_iA_i^\dagger) U_n^\dagger\rangle_\rho\right|\\
=&\sum\limits_{m\in \gamma_l,n\in \gamma_{l'},l\neq l'}\left|\langle U_m(\bigotimes\limits_{i=1}^NA_iA_i^\dagger) U_n^\dagger\rangle_\rho|+\sum\limits_{m,n\in \gamma_{k+1},m\neq n}|\langle U_m(\bigotimes\limits_{i=1}^NA_iA_i^\dagger) U_n^\dagger\rangle_\rho\right|\\
\leq&\sum\limits_{m\in \gamma_l,n\in \gamma_{l'},l\neq l'}\sqrt{\langle(\bigotimes\limits_{i=1}^NA_iA_i^\dagger)\rangle_\rho\langle U_mU_n(\bigotimes\limits_{i=1}^NA_iA_i^\dagger)U_n^\dagger U_m^\dagger\rangle_\rho}\\
&+\dfrac{1}{2}\sum\limits_{m,n\in \gamma_l,m\neq n}(\langle U_m(\bigotimes\limits_{i=1}^NA_iA_i^\dagger) U_m^\dagger\rangle_\rho+\langle U_n(\bigotimes\limits_{i=1}^NA_iA_i^\dagger) U_n^\dagger\rangle_\rho)\\
\leq&\sum\limits_{m\neq n}\sqrt{\langle \bigotimes\limits_{i=1}^NA_iA_i^\dagger\rangle_\rho\langle U_mU_n(\bigotimes\limits_{i=1}^nA_iA_i^\dagger)U_n^\dagger U_m^\dagger\rangle_\rho}+(N-k-1)\sum\limits_m\langle U_m(\bigotimes\limits_{i=1}^NA_iA_i^\dagger) U_m^\dagger\rangle_\rho.
\end{array}$$
Hence, inequality  (\ref{k-pro-s}) holds for any  pure state containing  $k$ unentangled particles. It is easy to prove that it is also right for any mixed state containing at least $k$ unentangled particles by utilizing absolute value inequality, inequality (\ref{k-pro-s}) for pure states, and the Cauchy-Schwarz inequality (\ref{CSI2}).

\textbf{Theorem 4.} For any $N$-partite fully separable state $\rho$, one has
\begin{equation}\label{f}
\begin{array}{rl}
\left|\langle U_m(\bigotimes\limits_{i=1}^NA_iA_i^\dagger) U_n^\dagger\rangle_\rho\right|
\leq\sqrt{\langle(\bigotimes\limits_{i=1}^NA_iA_i^\dagger)\rangle_\rho\langle U_mU_n(\bigotimes\limits_{i=1}^NA_iA_i^\dagger)U_n^\dagger U_m^\dagger\rangle_\rho}
\end{array}
\end{equation}
for any $m\neq n$.  If $\rho$ does not satisfy the above inequality (\ref{f}), then it is entangled.

\emph{Proof.} The proof of this result is quite similar to Theorem 3. Note that there is only one case that
$m,n$  belong to different $\gamma_l$ if $\rho=|\psi\rangle\langle\psi|$ is fully separable pure state, which ensures that inequality (\ref{f}) is true for fully separable pure state. Hence, inequality (\ref{f}) also holds for fully separable mixed states.

The following examples shows that the power of our results by comparison with Observation 5 in Ref.\cite{Goth2012}.

\textbf{Example 1.} For the family of  quantum  states
$$\rho(p)=p|\Psi_{5}\rangle\langle \Psi_{5}|+\dfrac{1-p}{5^{5}}\textbf{1},$$
where $|\Psi_{5}\rangle=\dfrac{1}{\sqrt{5}}\sum\limits_{i=0}^{4}|iiiii\rangle$.

Applying Theorem 1 by choosing $A_i=|1\rangle\langle0|, B_i=|0\rangle\langle0|$, we can get that,
if $p>0.0016$, $\rho$ contains at most $3$ unentangled particles;
if $p>0.0173$, $\rho$ contains at most $2$ unentangled particles;
 if $p>0.0325$, $\rho$ contains at most $1$ unentangled particles;
if $p>0.0399$, $\rho$ contains at most $0$ unentangled particles. But  Observation 5 in Ref.\cite{Goth2012} cannot detect any quantum states containing at most $k$ unentangled particles for $0\leq k\leq3$.

\textbf{Example 2.}  Consider the $N$-qubit mixed states,
\begin{eqnarray*}
\rho(p)=p|G\rangle\langle G|+\dfrac{1-p}{2^{N}}\textbf{1},
\end{eqnarray*}
where $|G\rangle=\dfrac{1}{\sqrt{2}}(|0\rangle^{\otimes N}+|1\rangle^{\otimes N}).$

Let $A_i=|1\rangle\langle0|, B_i=|0\rangle\langle0|$, then by using Theorem 1, we know that $\rho(p)$ contains at most $N-3$ unentangled particles when $p_{N-2}<p\leq1$, while by Observation 5 in Ref.\cite{Goth2012}, $\rho(p)$ contains at most $N-3$ unentangled particles when $p'_{N-2}<p\leq1$. The exact value of $p_{N-2}$ and $p'_{N-2}$ for $N$ = 9,10,\ldots,15 are shown in the Table I.

\begin{table}
\caption{\label{tab:table1}
For $\rho(p)=p|G\rangle\langle G|+\dfrac{1-p}{2^{N}}\textbf{1}$, the thresholds of $p_{N-2}$, $p'_{N-2}$  for  the quantum states containing at most $N-3$ unentangled particles  detected by Theorem 1 and
Observation 5 in Ref.\cite{Goth2012} for $9\leq N\leq15$, respectively, are illustrated.
When $p_{N-2}<p\leq1$ and $p'_{N-2}<p\leq1$, $\rho(p)$ contains at most $N-3$ unentangled particles by Theorem 1 and  Observation 5 in Ref.\cite{Goth2012}, respectively. Clearly,
Theorem 1 can detect more  states containing at most $N-3$ unentangled particles than Observation 5 in Ref.\cite{Goth2012} for $9\leq N\leq15$.
}
\begin{ruledtabular}
\begin{tabular}{ccccccccc}
$N$&9&10&11&12&13&14&15\\
\hline
&&&&&&&&\\
$p_{N-2}$ &   0.1263 & 0.0824 & 0.0519 &
0.0317 & 0.0189 & 0.0111 & 0.0064\\
&&&&&&&&\\
$p_{N-2}'$ & 0.1547 & 0.1350 & 0.1197 & 0.1076 & 0.0977 & 0.0894 & 0.0824
\end{tabular}
\end{ruledtabular}
\end{table}
\textbf{Example 3.}  Consider the $N$-qubit mixed states,
$$\rho(p,q)=p|W_{N}\rangle\langle W_{N}|+q\sigma_x^{\otimes N}|W_{N}\rangle\langle W_{N}|\sigma_x^{\otimes N}+\dfrac{1-p-q}{2^{N}}\textbf{1}.$$
Here $|W_{N}\rangle=\dfrac{1}{\sqrt{N}}(|10\cdots0\rangle+|01\cdots0\rangle+\cdots+|0\cdots01\rangle)$
and $\sigma_x$ is the Pauli matrix.

By choosing $u_m=\sigma_x$, $A_i=|0\rangle\langle0|$ (or $A_i=|1\rangle\langle0|$), our Theorem 3
can identify quantum states containing at most $k-1$ unentangled particles.  For $k=2$, the detection parameter spaces in which the quantum states contains at most 1 unentangled particles when $N = 6,7,8,9$ are shown in Fig. 1.
\begin{figure}[htbp]
\centering
\scalebox{0.8}[0.8]{\includegraphics{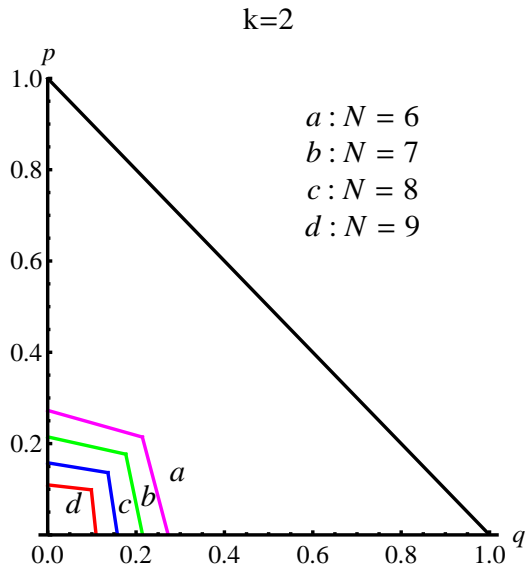}} \caption{Detection quality of  Theorem 3 for the state
$\rho(p,q)=p|W_{N}\rangle\langle W_{N}|+q\sigma_x^{\otimes N}|W_{N}\rangle\langle W_{N}|\sigma_x^{\otimes N}+\dfrac{1-p-q}{2^{N}}\textbf{1}$ for $k=2$ when $N=6,7,8,9$.  The area enclosed by
magenta $a$ (green line $b$, blue line $c$,  red line $d$), $p$ axis, line $q = 1- p$ and $q$ axis corresponds to the quantum states containing at most 1 unentangled particles when $N=6$ ($N=7$, $N=8$, $N=9$), respectively. }
\end{figure}

When $p=0$, the quantum state $\rho(p,q)$ is
$$\rho(q)=q\sigma_x^{\otimes N}|W_{N}\rangle\langle W_{N}|\sigma_x^{\otimes N}+\dfrac{1-q}{2^{N}}\textbf{1}.$$
By choosing $u_m=\sigma_x$,  $A_i=|1\rangle\langle0|$, our Theorem 3
can identify more quantum states containing at most $k-1$ unentangled particles than Observation 5 in Ref.\cite{Goth2012} when $N=8$.
For $1\leq k\leq6$, when $q_k<q\leq q'_k$, these quantum states containing at most $k-1$ unentangled particles which only can be detected by our
Theorem 3, but not by  Observation 5 in Ref.\cite{Goth2012}.  The exact value of $q_k$ and $q'_k$ for $1\leq k\leq6$ are shown in the Table II.

\begin{table}
\caption{\label{tab:table1}
For $\rho(q)=q\sigma_x^{\otimes N}|W_{N}\rangle\langle W_{N}|\sigma_x^{\otimes N}+\dfrac{1-q}{2^{N}}\textbf{1}$ when $N=8$, the thresholds of $q_k$, $q'_k$  for  the quantum states containing at most $k-1$ unentangled particles  detected by Theorem 3 and
Observation 5 in Ref.\cite{Goth2012} for $1\leq k\leq6$, respectively, are illustrated.
When $q_k<q\leq1$ and $q'_k<q\leq1$, $\rho(q)$ contains at most $k-1$ unentangled particles detected by our Theorem 3 and
Observation 5 in Ref.\cite{Goth2012}, respectively.
 The symbol $\setminus$ means that Observation 5 in Ref.\cite{Goth2012} cannot identify any quantum states containing at most 0 unentangled particles and at most 1 unentangled particles.
}
\begin{ruledtabular}
\begin{tabular}{cccccccc}
$k$&1&2&3&4&5&6\\
\hline
&&&&&&&\\
$q_k$ &  0.2889 & 0.1579 & 0.1028 &
0.0725 & 0.0533 & 0.0400 \\
&&&&&&&\\
$q_k'$ & $\setminus$ &$\setminus$ & 0.8647 & 0.6392 & 0.4587 & 0.3234
\end{tabular}
\end{ruledtabular}
\end{table}

\section{Conclusion}
In this paper, we have investigated the problem of detection of  quantum states containing at most $k-1$ unentangled particles. Several criteria for detecting  states containing at most $k-1$ unentangled particles were presented for arbitrary dimensional  multipartite quantum systems. It turned out that our results were effective by some specific examples. We hope that our results can contribute to a further understanding of entanglement properties of multipartite quantum systems.

\begin{acknowledgments}
This work was supported by the National Natural Science Foundation of China under Grant Nos: 12071110, 11701135 and 11947073; the Hebei Natural Science Foundation of China under Grant Nos: A2020205014, A2018205125 and A2017403025, and the Education Department of Hebei Province Natural Science Foundation under Grant No: ZD2020167, and the Foundation of  Hebei GEO University under Grant No. BQ201615.
\end{acknowledgments}

\end{document}